# Hybrid-functional and quasi-particle calculations of band structures of $Mg_2Si$, $Mg_2Ge$, and $Mg_2Sn$

Byungki Ryu,[1,a)] Sungjin Park,[1] Eun-Ae Choi,[2] Johannes de Boor,[3] Pawel Ziolkowski,[3] Jaywan Chung,[1] and SuDong Park[1]

[1]*Energy Conversion Research Center, Electrical Materials Research Division, Korea Electrotechnology Research Institute (KERI), Changwon 51543, Republic of Korea*

[2]*Computational Materials Department, Materials Processing Innovation Research Division, Korea Institute of Materials Science (KIMS), Changwon 51508, Republic of Korea*

3 *German Aerospace Center (DLR) - Institute of Materials Research, Cologne 51147, Germany*

We perform hybrid functional and quasi-particle band structure calculations with spin–orbit interaction to investigate the band structures of $Mg_2Si$, $Mg_2Ge$, and $Mg_2Sn$. For all $Mg_2X$ materials, where X = Si, Ge, and Sn, the characteristics of band edge states, i.e., band and valley degeneracies, and orbital characters, are found to be conserved, independent of the computational schemes such as density functional generalized gradient approximation, hybrid functionals, or quasi-particle calculations. However, the magnitude of the calculated band gap varies significantly with the computational schemes. Within density-functional calculations, the one-particle band gaps of $Mg_2Si$, $Mg_2Ge$, and $Mg_2Sn$ are 0.191, 0.090, and -0.346 eV, respectively, and thus severely underestimated compared to the experimental gaps, due to the band gap error in the density functional theory and the significant relativistic effect on the low-energy band structures. By employing hybrid-functional calculations with a 35% fraction of the exact Hartree–Fock exchange energy (HSE-35%), we overcame the negative band gap issue in $Mg_2Sn$. Finally, in quasi-particle calculations on top of the HSE-35% Hamiltonians, we obtained band gaps of 0.835, 0.759, and 0.244 eV for $Mg_2Si$, $Mg_2Ge$, and $Mg_2Sn$, respectively, consistent with the experimental band gaps of 0.77, 0.74, and 0.36 eV, respectively.

________________________________

a) Corresponding author: byungkiryu@keri.re.kr.

## I. INTRODUCTION

**Thermoelectric technology enables eco-friendly energy harvesting by converting unused or waste heat into electricity** [1,2]. For the fast industrialization of thermoelectric technology, highly efficient materials and modules are demanded strongly. As thermoelectric efficiency can be approximately determined by the dimensionless thermoelectric figures of merit $ZT$ [1-4], high $ZT$ materials have been explored in which $ZT$ is expressed as $ZT = (\alpha^2\sigma/\kappa)T$, where $\alpha$, $\sigma$, $\kappa$, and $T$ are the Seebeck coefficient, electrical conductivity, thermal conductivity, and absolute temperature, respectively. From the relation between electrical and thermal transport and $ZT$, a large power factor (PF) $\alpha^2\sigma$ and a low thermal conductivity $\kappa$ are important to obtain a high performance thermoelectric system [5,6]. Several material strategies have been developed to increase the PF: Fermi-level tuning [5-10], band convergence for charge transport [11,12], formation of resonant states [13,14], and energy filtering of charge carriers [15-22]. To decrease the thermal conductivity by phonon–alloy scattering in solid solutions [2,23], intrinsic and extrinsic nanosized inclusions or precipitations [24,25], nanostructuring due to smaller grain sizes [26,27], and intrinsically strong anharmonic phonon scattering [28,29] have been introduced.

**Mg-based binary silicides, i.e., $Mg_2X$ (X = Si, Ge, Sn) compounds and their alloys are eco-friendly thermoelectric materials, as they consist of nontoxic and earth-abundant materials, unlike tellurides or other heavy-element-based compounds.** Mg-based silicides crystallize in the antifluorite structure with a chemical formula of $Mg_2Si$ [30]. The Si atoms form an fcc sublattice with lattice parameter $a$, while the Mg atoms form a simple cubic sublattice with lattice parameter $a/2$. By mixing $Mg_2Si$ with $Mg_2Ge$ or $Mg_2Sn$, $Mg_2Si$-based alloys can exhibit a relatively high n-type $ZT$ exceeding 1 at a temperature from 600 to 800 K [12, 31-35], with a high power factor and low thermal conductivity, compared to the binary $Mg_2Si$ [36].

**$Mg_2X$-based alloys exhibit low thermal conductivity, owing to their complex microstructures.** In these materials, thermal energy is primarily transferred by phonons that are bosons. Thus, all phonon modes contribute to thermal conduction [37]. Furthermore, $Mg_2(Si,Sn)$ alloys exhibit complex microstructures owing to the miscibility gap, thus resulting in the phase separation to Si- and Sn-rich $Mg_2(Si,Sn)$ [33, 34] . In this alloy, each phase can serve as a point-disorder region because it is a disordered solid solution, while the interface between Si- and Sn-rich regions can block the phonon transport. Thus, in $Mg_2Si$-related silicide alloys, the naturally formed superlattice-like structure is reported to be responsible for its low thermal conductivity of approximately 1–3 W/m/K [34,35]. Similarly, our recent work revealed Sb-induced microstructure inhomogeneities in $Mg_2(Si,Sn)$ alloys [36].

**$Mg_2X$ compounds exhibit a high thermoelectric PF [12,21,35], owing to a unique n-type electronic structure [45,46].** The experimental band gaps of $Mg_2Si$, $Mg_2Ge$, and $Mg_2Sn$ are known to be 0.77, 0.74, and 0.36 eV, respectively [1] [47-50]. The valence and conduction band edge states are located at Γ and X points, respectively [45-47]. For $Mg_2Si$ and $Mg_2Ge$, the conduction band minimum (CBM) state (X1) does not belong to Mg or Si atoms, while the valence band maximum states are localized near atomic sites. Instead, the X1 state is localized near the interstitial site [45]. Under the tensile strain on $Mg_2Si$ and $Mg_2Ge$, this CBM state ascends, while the next CBM state (X3), composed of Mg s states, descends, resulting in the band convergence between X1 and X3 [46]. Note that the lattice parameter varies when $Mg_2Si$ is alloyed with $Mg_2Sn$. Thus, as discussed in [46], in the Mg$_2$(Si,Sn) solid solution, the lattice parameter variation and the conduction band convergence are responsible for the high n-type PF in Mg$_2$(Si,Sn) solid solutions. Furthermore, Mg$_2$X compounds exhibit a large density-of-states effective mass with a small transport effective mass because of their anisotropic band structures [51,52]. Meanwhile, creating p-type Mg$_2$X-based compounds and alloys by impurity doping such as Ag- and Li doping has been challenging [53-56]. However, the p-type PFs for Mg$_2$X and their alloys are found to be smaller than those of n-types [21,57], owing to the lower p-doping efficiency and a smaller band degeneracy for VBM, compared to the n-type[57].

**The recent theoretical and computational investigations on quantum transport have revealed the importance of nanostructuring on charge and heat transport.** Nanostructuring can be effective in increasing electrical conductivity and PF, in addition to suppressing bipolar thermal conductivity, especially for bipolar or narrow gap materials with interface-rich structures, and this results in enhanced thermoelectric performance [20-22]. However, to optimize thermoelectric materials for better thermoelectric performance, it is essential to correlate experimental results with theoretical studies. Thus, it is important to understand the atomic and electronic structures of Mg$_2$X compounds and their alloys thoroughly.

**The electronic structure of Mg$_2$X compounds has been investigated for the further understanding of Mg$_2$Si-related alloy systems** using first-principles density-functional theory (DFT) pseudopotential calculations [58,59]. However, because of the band gap error in DFT [60], the computed band gaps for Mg$_2$X were still highly underestimated compared to the experimental findings. To overcome the DFT band gap error, hybrid-functional calculations were adopted for Mg$_2$Si and Mg$_2$Sn [46]. However, the band gap is still small in hybrid-functional calculations. Quasi-particle GW calculations were conducted for $Mg_2Si$ and $Mg_2Ge$ [61], without the relativistic effect that is crucial for heavy elements such as Ge and Sn [52]. However, in 2018, Shi and Kioupakis reported relativistic quasi-particle GW band gaps with high accuracy for $Mg_2Si$, $Mg_2Ge$, and $Mg_2Sn$ [51]. Furthermore, in GW calculations, the proper description of initial wavefunctions and eigenvalues are important. In this context, Shi and Kioupakis corrected the wrong band occupation number. Additionally, the correction owing to spin–orbit interactions (SOIs) were included on top of the GW band structures in a non-self-consistent manner [62,63].

**The band gap is critical for thermoelectric transport, especially at higher temperatures where the thermal energy can overcome the band gap energy or the band offset potentials between materials.** For example, a small gap results in a smaller Seebeck coefficient and a larger value of thermal conductivity owing to enhanced bipolar conduction [64]. Moreover, the band gap size and position of band edges are crucial because they are correlated to various transport properties such as dopant ionization energy and dopability [65], band offset potential for carriers [66,67], and work function differences between materials responsible for carrier filtering [68]. However, the $Mg_2Sn$ band gap is still 0.142 eV [51] in state-of-art relativistic GW calculations, i.e., smaller by approximately 0.1 to 0.2 eV than the experimentally reported values. Although Shi and Kioupakis corrected the band occupation, an error may exist in the starting wavefunction owing to the overlap between the valence and conduction bands. Meanwhile, it was found that, in small-gap materials, one-shot GW calculations can be erroneous if the starting wavefunctions are wrong [69,70].

**In this study, to overcome the band gap underestimation owing to erroneous wavefunctions and wrong band occupations, we perform hybrid functional and quasi-**

**particle calculations of the band structures of Mg$_2$X, where X = Si, Ge, and Sn.** Using hybrid-functional calculations, we overcome the negative band gap problem in Mg$_2$Sn and correct the incorrect band occupation near the negative band gap. Using the one-shot GW calculations on top of the hybrid-functionals, we finally obtain reliable band gap values for Mg$_2$X, which are consistent to the measured values.

## II. CALCULATION METHOD

**We performed the DFT [71,72], hybrid-functional [73], and quasi-particle GW calculations [74-76] to calculate the atomic and electronic structures of Mg$_2$X, where X = Si, Ge, and Sn.** For DFT calculations, we used the projector-augmented-wave (PAW) pseudopotential [77], generalized-gradient-approximation parameterized by Perdew–Burke–Ernzerhof (PBE) [78], which were implemented in the VASP planewave code [79,80]: for the atomic potentials, we used the Mg potential of 'PAW Mg_pv_GW 20Apr2010', Si potential of 'PAW Si_GW 19Mar2012', Ge potential of 'PAW Ge_d_GW 17Dec2007', and Sn potential of 'PAW_PBE Sn_d_GW 20Mar2012', which are implemented in VASP code. The lattice parameters were optimized within DFT-PBE, without SOIs. The optimal lattice parameters from DFT-PBE calculations were 6.348, 6.410, and 6.801 Å for Mg$_2$Si, Mg$_2$Ge, and Mg$_2$Sn, respectively. Note that the calculated lattice parameters are very close to the experimental lattice parameters, 6.338, 6.384, and 6.750 Å for Mg$_2$Si, Mg$_2$Ge, and Mg$_2$Sn, respectively [48,81].

For the band structure calculations of the primitive cell of Mg$_2$X containing two Mg and one X atoms, we used the 6 × 6 × 6 MP k-point mesh [82]. For hybrid-functional DFT calculations, we used the Heyd–Scuseria–Ernzerhof functionals (HSE06) with mixing parameter of 25, 30, and 35% for the Hartree–Fock exact exchange with a screening parameter of 0.2 Å$^{-1}$ [73]. For the GW calculations, we used DFT-PBE or hybrid-DFT Hamiltonians for starting electronic wavefunctions and the Kohn–Sham eigenvalues. Subsequently, the self-energy was computed from G and W, and the quasi-particle band structure was calculated. In GW calculations, the number unoccupied bands are important to obtain the reliable converged band gap. Here, in total, we used the 240, 360, and 360 bands while there are 10, 10, and 15 occupied bands for Mg$_2$Si, Mg$_2$Ge, and Mg$_2$Sn, respectively. For GW with SOI calculations, the numbers of bands are doubled.

For the GW on top of the PBE Hamiltonian, we used four different schemes for the GW calculations: one-shot $G_0W_0$ approximation with one-shot G and W; $GW_0$ approximation with updated G with fixed W; self-consistent $GW_0$ (sc$GW_0$) including an off-diagonal component with updated G and fixed W; fully self-consistent GW (scGW), where G and W are updated. For the GW on top of the HSE Hamiltonian, we only considered the $G_0W_0$ calculations with HSE-25, HSE-30, and HSE-35. To observe the effect of SOI on the band structure, we performed the GW calculations with and without SOI.

## III. RESULTS AND DISCUSSION

**Figure 1 shows the one-particle DFT-PBE band structures for Mg$_2$X compounds (X = Si, Ge, Sn), without and with SOI.** Mg$_2$X binary compounds exhibit highly similar band dispersions especially for the valence bands. For all Mg$_2$X, the p-orbital bands of the X atoms are located near the energy range from –5 to 0 eV, with respect to the VBM energy.

Three VBM band states are degenerated at the Γ point if SOI effects are not taken into account. The CBM states of $Mg_2X$ are found at point X with a valley degeneracy of three, thus exhibiting an indirect band gap nature. With SOI, the three VBM bands the Γ point are split into 4 and 2 bands at, while the CBM bands at the L point are not. Note that, as going from X = Si to Sn, the spin-orbit splitting becomes larger. Table 1 shows the characteristics of the band edge states for $Mg_2X$. The $Mg_2X$ compounds exhibit different CBM states: for $Mg_2Si$ and $Mg_2Ge$, the CBM is the X1 state localized at the interstitial site, and the next CBM state (C+1) is the X3 state localized at the Mg s orbitals (see Figure 2). However, for $Mg_2Sn$, the ordering between X1 and X3 is inverted: X3 is the CBM, while X1 is the C+1 state. It was reported that the band inversion of $Mg_2X$ was related to the lattice strain [46]. Similarly, in Table 1, we summarize the effect of volume expansion on the band ordering of CBM and C+1.

**In DFT-PBE calculations, the band gaps are severely underestimated,** i.e., 0.202, 0.151, and -0.190 eV, for $Mg_2Si$, $Mg_2Ge$, and $Mg_2Sn$, respectively. With SOI effect, the band gaps are reduced to 0.191, 0.090, and –0.346 eV, respectively (see Figure 1). It is noteworthy that the SOI affects the band gap significantly, consistent to a previous report [51]. Furthermore, $Mg_2Sn$ has a negative band gap energy. To overcome the band gap underestimation of $Mg_2X$, we employed hybrid-functional and quasi-particle GW calculations.

**Table 2 shows the positions of band edges and the indirect and direct band gaps of $Mg_2Si$ using DFT-PBE, hybrid-functional, and quasi-particle GW calculations.** The band edges are computed with respect to the average of the electrostatic local potential over the unit cell, which can be used for band alignment using the reference potential method [66-68, 83]. It is clearly observed that the hybrid-functional calculations exhibit larger band gaps than DFT-PBE; however, the results from HSE-25, HSE-30, and HSE-35 are still smaller than the experimental values. To overcome the band gap underestimation of DFT-PBE and HSEs, we performed quasi-particle GW calculations to obtain a quasi-particle band gap. We found that the one-shot approximation still underestimated the band gap, even though SOI was not included. However, by iterating G or W, we finally increased the band gap to a level comparable to the experimental value (~0.80 eV). This implies that the initial wavefunctions might be inexact. However, the scGW computations are extremely time consuming. To minimize the computation time and use the corrected starting wavefunctions and Kohn–Sham eigenvalues, we calculated the quasi-particle band gap using the one-shot $G_0W_0$ calculations on top of HSE-25, HSE-30, and HSE-35, namely $G_0W_0$ (HSE-25), $G_0W_0$ (HSE-30), and $G_0W_0$ (HSE-35), respectively. Finally, we obtained a band gap of 0.835 eV using $G_0W_0$ (HSE-35), which was slightly larger than the experimental gap of 0.77 eV.

**Table 3 shows the positions of band edges and the indirect and direct band gaps of Mg$_2$Ge using various computational schemes.** We observed that the band gaps were underestimated even for the HSE-35 and one-shot GW calculations. By iterating G and W, the self-consistent quasi-particle band gap (0.768 eV) becomes comparable to the experimental gap of 0.74 eV. It is noteworthy that the inclusion of SOI is important for the band gap prediction of Mg$_2$Ge. In DFT-PBE, the SOI reduce the Mg$_2$Ge band gap by 0.06 eV. If we consider the SOI correction on the band gap, the self-consistent quasi-particle band gap might be smaller than 0.7 eV. This implies that the self-consistent band gap will be smaller than the experimental gap, owing to poor starting wavefunctions for Mg$_2$Ge. Therefore, by conducting quasi-particle GW calculations on top of the HSE-35, we corrected the initial wavefunctoins compared to density-functoinal calculations, and finally obtained the band gap of 0.759 eV with relativistic effects.

**Table 4 shows the band structure information for Mg$_2$Sn from DFT-PBE, HSE methods, and quasi-particle GW calculations.** In contrast to Mg$_2$Si and Mg$_2$Ge, Mg$_2$Sn exhibits a negative band gap in DFT-PBE. It is noteworthy that the band gap is still negative for HSE-25 with SOI. When we increased the fraction of the exact Hartree–Fock exchange to 30–35%, we finally obtained positive band gaps for Mg$_2$Sn. The band gaps of DFT-PBE, HSE-25, HSE-30, and HSE-35 were computed to be -0.346, -0.030, 0.052, and 0.134 eV. It is well known that wrong starting wavefunctions or wrong band occupations result in the wrong quasi-particle band gaps [51,69,70]. Thus, we must correct the electronic wavefunction by iterating G or W when we perform the GW calculations. We found that scGW$_0$ and scGW can increase the band gap of Mg$_2$Sn. However, the self-consistent GW methods are time consuming. Therefore, we conducted the GW calculations on top of the hybrid-functional results. By employing the HSE-35 electronic wavefunction with a positive band gap, we obtained a plausible band gap of 0.244 eV, consistent with the experimental gap of 0.36 eV. It is noteworthy that our computed gaps for Mg$_2$Sn are larger than Shi and Kioupakis's band gap of 0.142 eV.

**Table 5 shows the effect of SOI on the band structure of Mg$_2$X.** In DFT-PBE calculations, the SOI effect reduces the band gaps of Mg$_2$Si, Mg$_2$Ge, and Mg$_2$Sn by 0.011, 0.062, and 0.156 eV, respectively. Also SOI affects VBM splittings as the mass of the involved X atoms become heavier [51]. For Mg$_2$Si, a negligible SOI splitting exists for the VBM and a negligible band gap narrowing. However, for Mg$_2$Sn, the spin splitting of VBM is enlarged to 0.485 eV for DFT-PBE, consistent to the semi-relativitistic Korringa-Kohn-Rostoker calculations [52]. It is noteworthy that the spin splitting is nearly unchanged with different computational schemes. This implies that the inclusion of SOI on top of GW calculations is valid [51], compared to our GW calculations on top of SOI calculations.

**We estimated the band edge positions of VBM, X1, and X3 for a ternary solid solution $Mg_2(Si_{(1-x)}Sn_x)$ and subsequently computed its band gap (see Figure 3(a) and (b)).** The band edge positions of binary $Mg_2X$ with respect to the average local potential in the bulk are calculated using the GW calculations on top of the HSE-35 Hamiltonian. By linearly interpolating the values for the end point binary compounds, i.e., $Mg_2Si$ and $Mg_2Sn$, we estimated the band edge positions of the ternary solid solution $Mg_2(Si,Sn)$. Note that our supercell approach on the band edge calculations also confirm the *nearly* linear relation between band edge energies and composition. As shown in Figure 3(a), the VBM and X1 states of $Mg_2Si$ ascends when alloying with $Mg_2Sn$, while the X3 state descends. Consequently, the X1 and X3 states are converged and the band gap becomes smaller. In DFT-PBE with SOI calculations, the X1 and X3 states converged at approximately x = 0.357, which is inconsistent with the experimental reports. In the quasi-particle GW calculations, the X1 and X3 conduction bands converged at x = 0.525. Similarly, as shown in **Figure 3(c) and (d)**, the conduction band convergence of $Mg_2(Ge_{1-y}Sn_y)$ is occurred at y = 0.60. It is noteworthy that the conduction band convergence between X1 and X3 is observed experimentally at approximately x = 0.6 for $Mg_2(Si_{1-x}Sn_x)$ and y=0.75 for $Mg_2(Ge_{1-y}Sn_y)$ [12, 84]. The remaining differences can be due to the temperature dependency band structures of $Mg_2X$: we computed the band structures at 0 K, while the experimental observation is at 600 to 800 K.

## IV. CONCLUSIONS

**In conclusion, the band structures of $Mg_2X$ (X = Si, Ge, and Sn) were investigated using DFT-PBE, hybrid-functionals, and quasi-particle GW calculations.** Severe band gap underestimations occurred owing to DFT band gap errors and strong spin–orbit interactions. By correcting the initial wavefunctions and Kohn–Sham eigenvalues from the hybrid-functional calculations, we obtained a reliable quasi-particle band gap for $Mg_2X$ that was consistent to the experimental ones. Finally, the quasi-particle band structures of $Mg_2X$ well predict the compositions of the conduction band convergences in $Mg_2(Si,Sn)$ and $Mg_2(Ge,Sn)$ solid solutions.

**ACKNOWLEDGMENTS**

This work was supported by the Energy Efficiency & Resource Core Technology Program of the Korea Institute of Energy Technology Evaluation and Planning (KETEP) granted from the Ministry of Trade, Industry & Energy (MOTIE), Republic of Korea (no. 20188550000290, no. 20172010000830, no. 20162000000910). It was also supported by the Korea Electrotechnology Research Institute (KERI) Primary Research Program through the National Research Council of Science and Technology (NST) funded by the Ministry of Science and ICT (MSIT) of the Republic of Korea (No. 19-12-N0101-22). EAC is supported by the Fundamental Research Program (POC3360) of the Korea Institute of Materials Science (KIMS). Also, one of the authors (JdB) is partially funded by the Deutsche Forschungsgemeinschaft (DFG, German Research Foundation)-project number 396709363.

# TABLEs

**Table 1** For given lattice parameters, real space characteristics of band edge states of $Mg_2Si$, $Mg_2Ge$, and $Mg_2Sn$ are described without and with strain. For VBM, CBM, and C+1 states, the main contributions are written.

| Material | $Mg_2Si$ | $Mg_2Ge$ | $Mg_2Sn$ | $Mg_2Si$, strained | $Mg_2Sn$, strained |
|---|---|---|---|---|---|
| a [Å] | 6.348 | 6.410 | 6.801 | 6.801 | 6.348 |
| VBM | Si p | Ge p | Sn p | Si p | Sn p |
| CBM | Interstitial (X1) | Interstitial (X1) | Mg s (X3) | Mg s (X3) | Interstitial (X1) |
| C+1 | Mg s (X3) | Mg s (X3) | Interstitial (X1) | Interstitial (X1) | Mg s (X3) |

**Table 2** Band edge energies and band gap of $Mg_2Si$ calculated from DFT-PBE, hybrid functionals, and quasi-particle GW calculations. The VBM and CBM energies respect to the average local potential, the indirect band gap ($E_g^{(i)}$), and the direct band gap ($E_g^{(d)}$) are computed with and without SOI.

| | $Mg_2Si$ | | | | | | | |
|---|---|---|---|---|---|---|---|---|
| | no SOI [eV] | | | | SOI [eV] | | | |
| | VBM | CBM | $E_g^{(i)}$ | $E_g^{(d)}$ | VBM | CBM | $E_g^{(i)}$ | $E_g^{(d)}$ |
| PBE | 3.276 | 3.478 | 0.202 | 1.883 | 3.287 | 3.478 | 0.191 | 1.872 |
| HSE-25 | 3.021 | 3.589 | 0.568 | 2.493 | 3.032 | 3.588 | 0.556 | 2.481 |
| HSE-30 | | | | | 2.977 | 3.611 | 0.635 | 2.609 |
| HSE-35 | | | | | 2.920 | 3.635 | 0.715 | 2.740 |
| $G_0W_0$(PBE) | 2.777 | 3.429 | 0.652 | 2.428 | | | | |
| $GW_0$(PBE) | 2.712 | 3.414 | 0.702 | 2.508 | | | | |
| $scGW_0$(PBE) | 2.689 | 3.444 | 0.756 | 2.579 | | | | |
| $G_0W_0$(HSE-25) | 2.704 | 3.521 | 0.817 | 2.685 | 2.718 | 3.481 | 0.763 | 2.625 |
| $G_0W_0$(HSE-30) | | | | | 2.702 | 3.500 | 0.798 | 2.679 |
| $G_0W_0$(HSE-35) | | | | | 2.685 | 3.520 | 0.835 | 2.734 |
| Experiment | | | | | | | 0.77 [48] | |

**Table 3** Band edge energies and band gap of Mg$_2$Ge calculated from DFT-PBE, hybrid functionals, and quasi-particle GW calculations. The VBM and CBM energies respect to the average local potential, the indirect band gap ($E_g^{(i)}$), and the direct band gap ($E_g^{(d)}$) are computed with and without SOI.

| | Mg$_2$Ge | | | | | | | |
| --- | --- | --- | --- | --- | --- | --- | --- | --- |
| | no SOI [eV] | | | | SOI [eV] | | | |
| | VBM | CBM | $E_g^{(i)}$ | $E_g^{(d)}$ | VBM | CBM | $E_g^{(i)}$ | $E_g^{(d)}$ |
| PBE | 2.540 | 2.692 | 0.151 | 0.965 | 2.602 | 2.691 | 0.090 | 0.904 |
| HSE-25 | 2.277 | 2.800 | 0.523 | 1.603 | 2.343 | 2.799 | 0.457 | 1.536 |
| HSE-30 | | | | | 2.287 | 2.822 | 0.536 | 1.671 |
| HSE-35 | | | | | 2.229 | 2.846 | 0.617 | 1.809 |
| G$_0$W$_0$(PBE) | 1.980 | 2.619 | 0.639 | 1.642 | | | | |
| GW$_0$(PBE) | 1.874 | 2.591 | 0.717 | 1.742 | | | | |
| scGW$_0$(PBE) | 1.852 | 2.620 | 0.768 | 1.805 | | | | |
| G$_0$W$_0$(HSE-25) | 1.920 | 2.719 | 0.799 | 1.910 | 1.987 | 2.677 | 0.690 | 1.801 |
| G$_0$W$_0$(HSE-30) | | | | | 1.973 | 2.697 | 0.724 | 1.858 |
| G$_0$W$_0$(HSE-35) | | | | | 1.959 | 2.718 | 0.759 | 1.916 |
| Experiment | | | | | | | 0.74 [48] | |

**Table 4** Band edge energies and band gap of Mg$_2$Sn calculated from DFT-PBE, hybrid functionals, and quasi-particle GW calculations. The VBM and CBM energies respect to the average local potential, the indirect band gap ($E_g^{(i)}$), and the direct band gap ($E_g^{(d)}$) are computed with and without SOI.

| | Mg$_2$Sn | | | | | | | |
| --- | --- | --- | --- | --- | --- | --- | --- | --- |
| | no SOI [eV] | | | | SOI [eV] | | | |
| | VBM | CBM | $E_g^{(i)}$ | $E_g^{(d)}$ | VBM | CBM | $E_g^{(i)}$ | $E_g^{(d)}$ |
| PBE | 3.603 | 3.413 | -0.190 | 1.312 | 3.759 | 3.413 | -0.346 | 1.161 |
| HSE-25 | 3.391 | 3.536 | 0.146 | 1.916 | 3.560 | 3.530 | -0.030 | 1.745 |
| HSE-30 | | | | | 3.510 | 3.562 | 0.052 | 1.878 |
| HSE-35 | | | | | 3.459 | 3.594 | 0.134 | 2.012 |
| G$_0$W$_0$(PBE) | 3.186 | 3.272 | 0.087 | 1.751 | | | | |
| GW$_0$(PBE) | 3.055 | 3.313 | 0.258 | 1.817 | | | | |
| scGW$_0$(PBE) | 3.030 | 3.354 | 0.324 | 1.967 | | | | |
| G$_0$W$_0$(HSE-25) | 3.066 | 3.447 | 0.381 | 2.092 | 3.287 | 3.364 | 0.077 | 1.827 |
| G$_0$W$_0$(HSE-30) | | | | | 3.237 | 3.414 | 0.177 | 1.925 |
| G$_0$W$_0$(HSE-35) | | | | | 3.205 | 3.449 | 0.244 | 1.999 |
| Experiment | | | | | | | 0.36 [48] | |

**Table 5 Effect of SOI on band gaps, band offset between CBM and C+1 states, and SOI splitting at VBM for Mg$_2$Si, Mg$_2$Ge, and Mg$_2$Sn.**

| Material | Inclusion of SOI | Computational scheme | Band gap [eV] | $E_{(C+1)} - E_{CBM}$ | SOI splitting at VBM |
|---|---|---|---|---|---|
| Mg$_2$Si | Without soi | PBE | 0.202 | 0.204 | |
| | With soi | PBE | 0.191 | 0.204 | 0.033 |
| | | HSE-35 | 0.715 | 0.339 | 0.337 |
| | | GW(HSE-35) | 0.835 | 0.338 | 0.033 |
| Mg$_2$Ge | Without soi | PBE | 0.151 | 0.363 | |
| | With soi | PBE | 0.090 | 0.364 | 0.190 |
| | | HSE-35 | 0.617 | 0.482 | 0.207 |
| | | GW(HSE-35) | 0.759 | 0.454 | 0.186 |
| Mg$_2$Sn | Without soi | PBE | -0.190 | -0.368 | |
| | With soi | PBE | -0.346 | -0.368 | 0.485 |
| | | HSE-35 | 0.134 | -0.335 | 0.518 |
| | | GW(HSE-35) | 0.244 | -0.306 | 0.464 |

**FIGURE CAPTIONS**

Fig.1. DFT-PBE band structures are drawn for (a) $Mg_2Si$, (b) $Mg_2Ge$, and (c) $Mg_2Sn$ without SOI. Also DFT-PBE band structures are drawn for (d) $Mg_2Si$, (e) $Mg_2Ge$, and (f) $Mg_2Sn$ with SOI. The VBM energies are set to zero. X1 and X3 states denote the CBM and C+1 states for $Mg_2X$, respectively.

Fig.2. Charge density distributions of (a) VBM, (b) X1, and (C) X3 states in $Mg_2Si$. The VBM state is localized around the Si atoms. The X1 state is the interstitial charge states surrounded by Si atoms. The X3 state is localized around the Mg site.

Fig.3. (a) The predicted quasi-particle (QP) band energies of VBM, X1, and X3 in the alloy of $Mg_2(Si_{(1-x)}Sn_x)$ solid solutions from linear interpolation. (b) The predicted indirect band gap was derived from the interpolated energies for VBM, X1, and X3 in the alloy of $Mg_2(Si_{(1-x)}Sn_x)$ solid solutions. In (c) and (d), the QP band energies and the indirect band gap is drawn for $Mg_2(Ge_{(1-y)}Sn_y)$ solid solution, respectively.

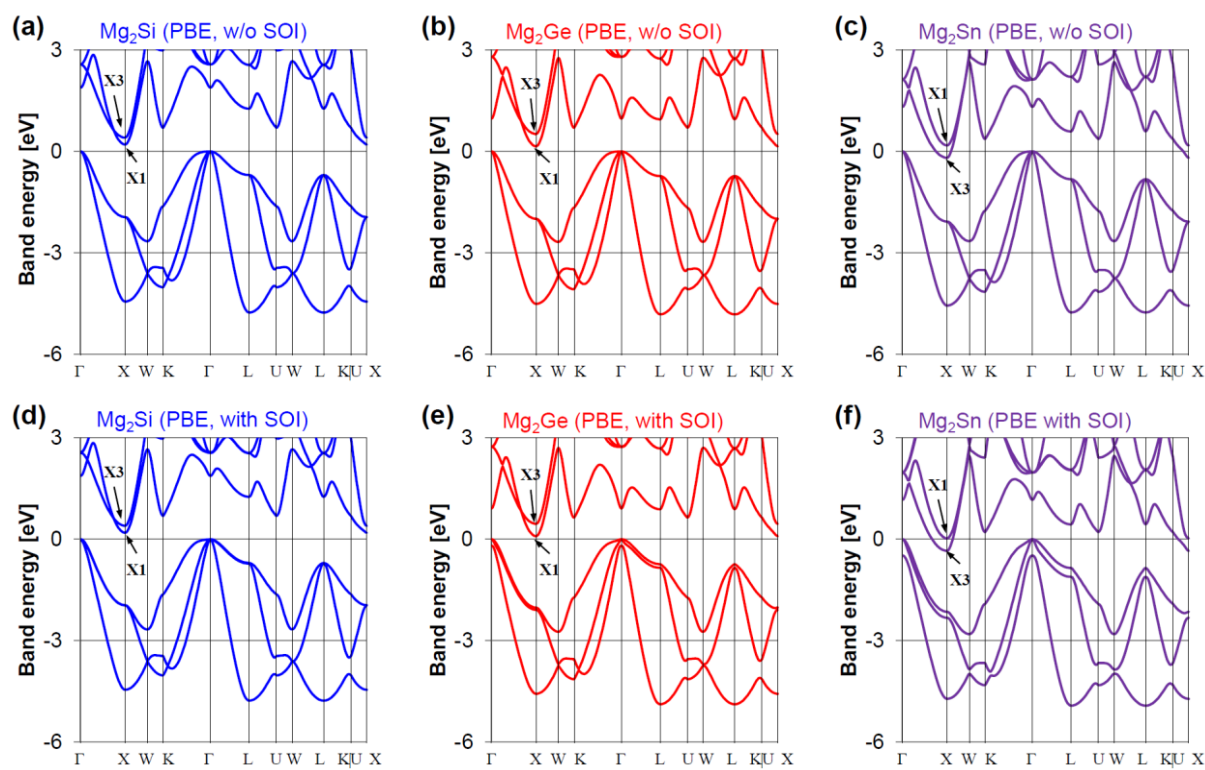

Fig.1. DFT-PBE band structures are drawn for (a) Mg$_2$Si, (b) Mg$_2$Ge, and (c) Mg$_2$Sn without SOI. Also DFT-PBE band structures are drawn for (d) Mg$_2$Si, (e) Mg$_2$Ge, and (f) Mg$_2$Sn with SOI. The VBM energies are set to zero. X1 and X3 states denote the CBM and C+1 states for Mg$_2$X, respectively.

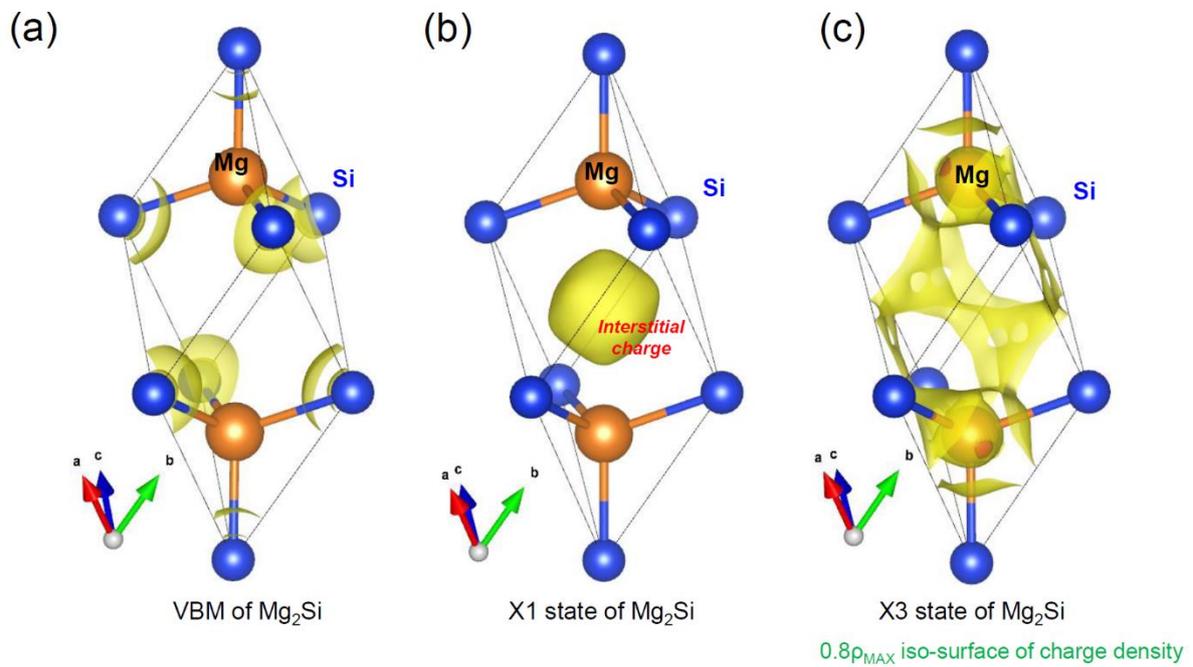

Fig.2. Charge density distributions of (a) VBM, (b) X1, and (C) X3 states in $Mg_2Si$. The VBM state is localized around the Si atoms. The X1 state is the interstitial charge states surrounded by Si atoms. The X3 state is localized around the Mg site.

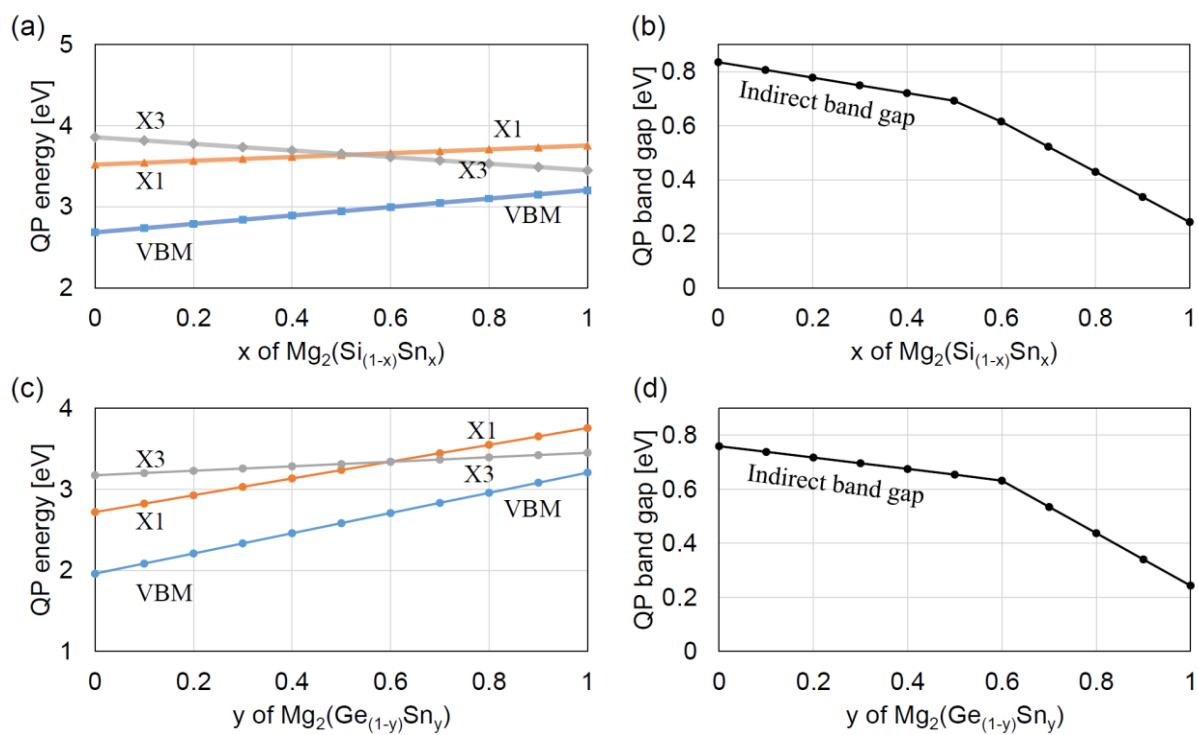

Fig.3. (a) The predicted quasi-particle (QP) band energies of VBM, X1, and X3 in the alloy of $Mg_2(Si_{(1-x)}Sn_x)$ solid solutions from linear interpolation. (b) The predicted indirect band gap was derived from the interpolated energies for VBM, X1, and X3 in the alloy of $Mg_2(Si_{(1-x)}Sn_x)$ solid solutions. In (c) and (d), the QP band energies and the indirect band gap is drawn for $Mg_2(Ge_{(1-y)}Sn_y)$ solid solution, respectively.